\documentclass[aps,pra,twocolumn,superscriptaddress]{revtex4}

\usepackage{color}
\usepackage{units}
\usepackage{graphicx}
\usepackage{amsmath}

\newcommand{\br}[1]{\left( #1 \right)}
\newcommand{\sqbr}[1]{\left[ #1 \right]}
\newcommand{\modbr}[1]{\left| #1 \right|}
\newcommand{\etal}{\emph{et al.~}}

\begin{document}

\title{Space-time coupling of shaped ultrafast ultraviolet pulses from an acousto-optic programmable dispersive filter}

\author{David J.\ McCabe}
\email{djmccabe1@googlemail.com}
\affiliation{ CNRS-Universit\'e de Toulouse,UPS, Laboratoire Collisions, Agr\'egats R\'eactivit\'e,
IRSAMC, F-31062 Toulouse, France}
\author{Dane R.\ Austin}
\affiliation{Clarendon Laboratory, Department of Physics, University of Oxford, Oxford, OX1 3PU, United Kingdom}
\author{Ayhan Tajalli}
\author{S\'{e}bastien Weber}
\author{Ian A.\ Walmsley}
\affiliation{Clarendon Laboratory, Department of Physics, University of Oxford, Oxford, OX1 3PU, United Kingdom}
\author{B\'{e}atrice Chatel}
\affiliation{ CNRS-Universit\'e de Toulouse,UPS, Laboratoire Collisions, Agr\'egats R\'eactivit\'e,
IRSAMC, F-31062 Toulouse, France}

\date{\today}

\begin{abstract}
A comprehensive experimental analysis of spatio-temporal coupling effects inherent to the acousto-optic programmable dispersive filter (AOPDF) is presented. Phase and amplitude measurements of the AOPDF transfer function are performed using spatially and spectrally resolved interferometry. Spatio-temporal and spatio-spectral coupling effects are presented for a range of shaped pulses that are commonly used in quantum control experiments. These effects are shown to be attributable to a single mechanism: a group-delay--dependent displacement of the shaped pulse. The physical mechanism is explained and excellent quantitative agreement between the measured and calculated coupling speed is obtained. The implications for quantum control experiments are discussed.
\end{abstract}

\maketitle

\section{Introduction}
\label{sec:intro}

Shaped and characterized femtosecond pulses are in widespread demand amongst the quantum control community \cite{Goswami2003,Ohmori2009,Dantus2004}. Through tailoring the phase, amplitude or polarization of the control pulse, the evolution of a quantum state may be manipulated in order to steer it towards a desired outcome. A typical scenario is the design of optical fields to control molecular motion, including the prospect of achieving site-specific chemistry and intramolecular rearrangements. During the last two decades, many impressive results \cite{Bonacic-Koutecky2006,Levis2001,Weinacht1999,Monmayrant2006} have arisen from technological breakthroughs in the generation of arbitrarily tailored pulses \cite{Monmayrant2010}.
 
Two principal active pulse-shaping techniques for ultrashort pulses are at the disposal of the experimentalist: a spatial light modulator (SLM) placed in the Fourier plane of a $4f$ zero-dispersion line \cite{Weiner2000,Monmayrant2004} or an acousto-optic programmable dispersive filter (AOPDF) \cite{Verluise2000}. Extensive studies of the $4f$ line have extended its available wavelength range and characterized its behaviour. In particular, it is now well known both experimentally and theoretically that such devices lead to spatio-temporal coupling effects, whereby the shaped electric field is dependent on the spatial position in the beam \cite{Danailov1989,Wefers1995,Wefers1996,Dorrer1998,Tanabe2002}. These studies have more recently been extended to the focal volume after a lens \cite{Sussman2008,Frei2009}.
 
By contrast, the AOPDF --- a newer technology within the control field --- has been less well characterized. Its first application entailed the corrective shaping of ultrashort pulses before an amplifier in order to improve compression of the amplified output \cite{Seres2003}. More recently, an angular dispersion effect which could affect such a laser chain has been presented \cite{Borzsonyi2010}. Further to this application, the AOPDF's shaping versatility, together with the large spectral range spanned (from the UV \cite{Coudreau2006} through the visible \cite{Monmayrant2005} to the near IR \cite{Verluise2000,Pittman2002}) renders it a valuable tool for control experiments \cite{Form2008}. In particular, the first--excited-state transitions of many organic and inorganic molecules lie in the UV wavelength range; hence the development of a practical UV pulse shaper is a great challenge and active field within the community. Very recently, interesting results have been obtained using shaped ultraviolet pulses \cite{Tseng2009, Roth2009,Greenfield2009}, and the variety of implementations of AOPDFs and SLMs is constantly increasing \cite{Monmayrant2010}. Amongst these contenders, the UV AOPDF based upon a KDP crystal is a good candidate, since it is versatile and tunable on a broad spectral range (\unit[250-410]{nm}) matching typical molecular electronic absorption bands \cite{Weber2010}. Nonetheless, to date no complete characterization has been performed of the spatio-temporal characteristics of AOPDF-shaped pulses --- in particular in the UV range. Indeed, some sources even assert AOPDFs to be entirely free of such effects \cite{Lee2009}, in contrast to the much maligned $4f$ line. At least one distortion, however, has already been identified: a lateral displacement which depends on the acoustic wave profile in the crystal \cite{Krebs2010}.

In this paper, we have undertaken the complete characterization of the space-time coupling effects produced by the AOPDF using spatially and spectrally resolved Fourier-transform interferometry (SSI) \cite{Monmayrant2010}. SSI is an interferometric technique that entails a relative measurement of the spectral phase between a reference and unknown pulse --- it thus lends itself to the measurement of the transfer function of a pulse shaper. As a metrology tool, SSI is suited to low pulse energies since it does not necessitate any nonlinear processes. (In the event that knowledge of the spectral phase of the input pulse \emph{per se} is required, absolute pulse characterization techniques may be applied \cite{Baum2004,Kane1994}.) This technique provides spatio-temporal resolution of the shaped pulses; it thus facilitates a comprehensive quantitative analysis of the ubiquitous spatio-temporal coupling induced by the AOPDF together with an explanation and numerical description of the physical mechanism. Our analysis encompasses a range of pulse shapes that are of the broadest utility to the control community.

\section{Methods}
\label{sec:methods}

\begin{figure}
\centering
\includegraphics[width = 0.9\columnwidth]{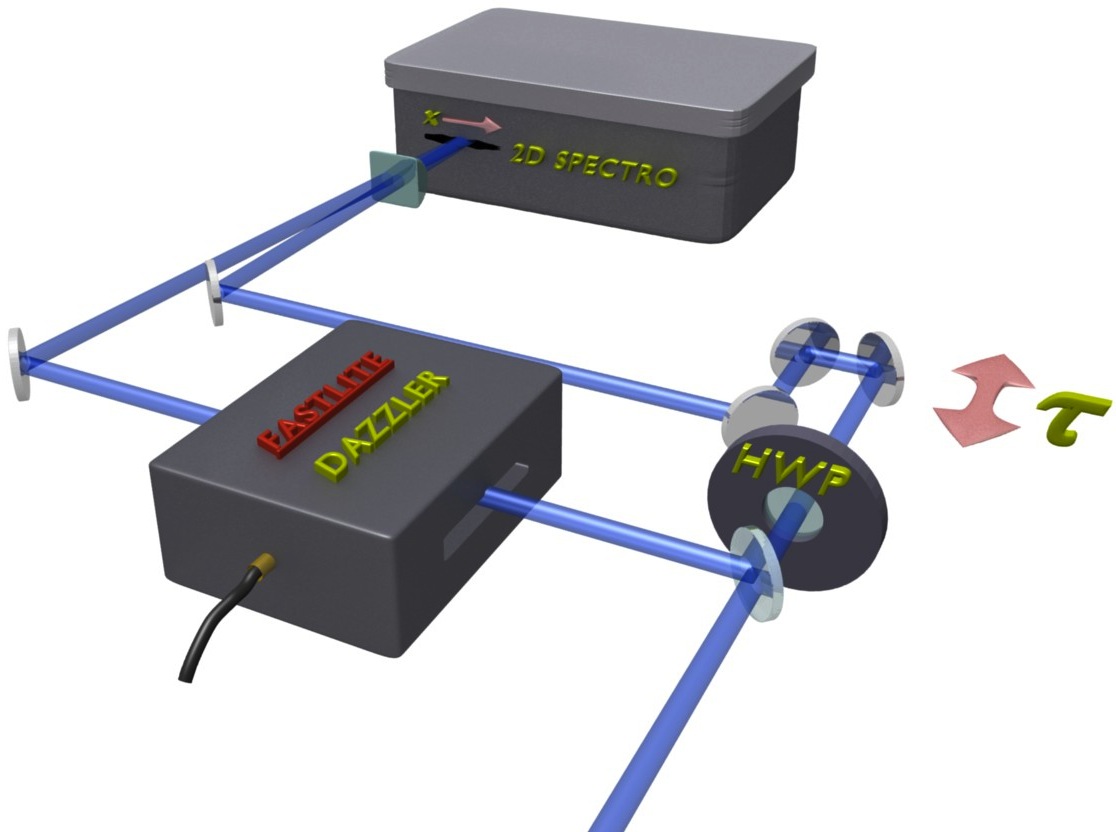}
\caption{The AOPDF (Fastlite Dazzler\texttrademark) SI characterization setup. The pulse shaper is placed in one arm of an interferometer. The unknown and reference arms are recombined at the entrance slit to a two-dimensional spectrometer with a slight angle and variable delay. The imaging spectrometer measures the resultant interference fringes, from which the relative spectral phase may be extracted. The spectrometer measures a spatially resolved spectrum along the slit axis $x$. A cylindrical lens focusses the beams onto the entrance slit of the spectrometer along the non-imaged spatial axis. A half-waveplate rotates the polarization in the reference beam arm.}
\label{fig:layout}
\end{figure}

The ultrafast source used for these experiments is an ultraviolet (UV) pulse train generated from a chirped-pulse amplified Ti:sapphire laser (CPA) \cite{Backus1998} via subsequent nonlinear interactions. The \unit[800]{nm} pulses are combined with their second harmonic at \unit[400]{nm} in order to generate the sum frequency at $\lambda_0 = \unit[267]{nm}$. Typical characteristics of the UV source are \unit[2]{$\mu$J} pulses with \unit[2]{nm} full-width at half-maximum (FWHM) bandwidth at the \unit[1]{kHz} repetition rate of the master laser. A typical beam width is around \unit[1-2]{mm}. Spatially resolved cross-correlation measurements of the UV pulses indicate a \unit[250]{fs} pulse duration without significant spatial wavefront distortion. (The pulse bandwidth would support a transform-limited duration of around \unit[50]{fs}; the difference is attributable to dispersive effects within the nonlinear crystals of the source.)

AOPDF pulse shapers are based on the dispersive propagation of light within an acousto-optic crystal. An incident ordinary optical wave interacts with a collinear acoustic wave, resulting in the diffraction of the optical wave onto the extraordinary axis. The spectral phase of a femtosecond optical pulse may be shaped via manipulation of the diffraction location for each spectral component along the length of the birefringent crystal; meanwhile the amplitude may be modulated via the size of the acoustic wave \cite{Kaplan2002}. A commercial AOPDF (the Fastlite Dazzler\texttrademark T-UV-260-410/T2), based on a \unit[75]{mm} KDP crystal designed for use at UV wavelengths, is employed for these experiments \cite{Coudreau2006,Weber2010}. The programmable temporal window is essentially fixed by the length of the crystal and the difference in refractive index of the crystal axes, and is about \unit[7]{ps} for this apparatus. A part of this window (for example, \unit[3]{ps} for a shaping window of three times the FWHM bandwidth according to the parameters given above) is required to self-compensate the natural dispersion induced by the KDP itself; if necessary this could be obviated by means of an external compressor.

The performance of the AOPDF is characterized using SSI \cite{Monmayrant2010,Tanabe2002}. The AOPDF is placed in one arm of an interferometer and its shaped output interferes with an unshaped reference arm (see Fig.\ \ref{fig:layout}). Since the AOPDF rotates polarization, a half-waveplate was placed in the reference beam arm. The two arms are combined with a small angle and a controllable relative delay at the entrance slit of a home-built imaging spectrometer \cite{Austin2009}. Since the spectrometer employs a two-dimensional detector, it is able to make a measurement of the spectrum as a function of position along the slit (aligned parallel with the plane of diffraction of the AOPDF). The detector is a charge-coupled device (CCD) camera (EHD Imaging UK-1158UV) with a pixel size of \unit[6.45]{$\mu$m}, and the spectrometer has an optical resolution of \unit[0.08]{nm} and \unit[40]{$\mu$m} along the spectral and spatial axes respectively. In order to increase signal, a cylindrical lens focusses the beams onto the entrance slit along the orthogonal spatial axis (i.e.\ along the non-imaged axis of the beam). The ensuing interferograms are detected with single-shot sensitivity.

\begin{figure}
\centering
\includegraphics [width = \columnwidth]{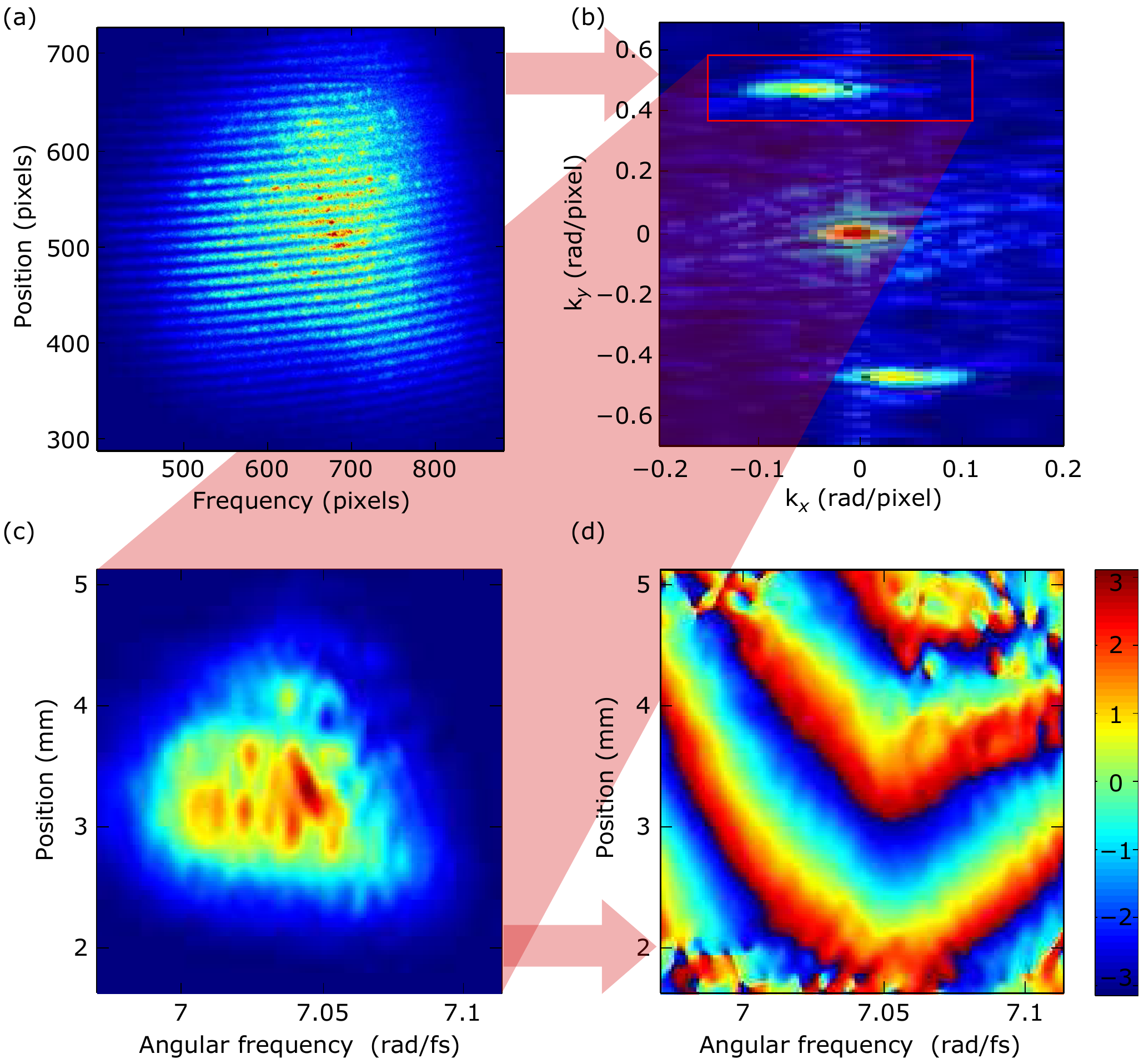}
\caption{The Fourier filtering process. (a) A raw interferogram measured by the spectrometer camera. (b) A two-dimensional Fourier transform is performed. An a.c.\ term is filtered out within the Fourier domain. (c) An inverse two-dimensional Fourier transform of this term isolates the final term of equation \ref{eq:SI}. The mapping onto calibrated frequency and position axes is calculated. (d) Extracted phase difference $\phi_\textrm{s}(x,\omega) - \phi_\textrm{r}(x,\omega)$, modulo $2\pi$. A subsequent procedure calibrates the camera pixels into physical units of frequency and position.}
\label{fig:data}
\end{figure}

The 2D interferogram measured by the spectrometer [see Fig.\ \ref{fig:data}(a)] is
\begin{align}
\label{eq:SI}
S(x,\omega) & = \modbr{A_\textrm{s}(x,\omega) e^{i\phi_\textrm{s}(x,\omega)} + A_\textrm{r}(x,\omega) e^{i\sqbr{\phi_\textrm{r}(x,\omega) + \omega \tau + k_x x}}}^2 \notag \\
& = \modbr{A_\textrm{s}(x,\omega)}^2 + \modbr{A_\textrm{r}(x,\omega)}^2 \notag \\ &+ \modbr{A_\textrm{s}(x,\omega)} \modbr{A_\textrm{r}(x,\omega)} \notag \\ & \times \cos \sqbr{\phi_\textrm{s}(x,\omega) - \phi_\textrm{r}(x,\omega) - \omega \tau - k_x x}.
\end{align}
Here $\tau$ is the time delay between the two pulses and $k_x$ is the difference between the transverse components of the propagation vectors (such that their subtended angle is $\theta = k_x/\modbr{\mathbf{k}}$). $A_\textrm{s}$, $A_\textrm{r}$, $\phi_\textrm{s}$ and $\phi_\textrm{r}$ denote the spatio-spectral amplitude and phase of the shaped (s) and reference (r) pulse respectively.

In order to extract a measurement of the spectral phase added by the AOPDF, this interferogram is Fourier transformed along both spatial and spectral dimensions. One of the a.c.\ terms is filtered out and inverse Fourier transformed with the carrier frequency removed. This isolates the final summand of equation \ref{eq:SI}, which contains the phase difference $\phi_\textrm{s}(x,\omega) - \phi_\textrm{r}(x,\omega)$ \cite{Takeda1982}. The spatial and spectral carriers, $k_x$ and $\tau$, are chosen in order to separate the a.c.\ and d.c.\ terms in the Fourier transform whilst ensuring that the fringe period is greater than the spectrometer resolution. In order to be able to handle complex temporal structure, a predominantly spatial carrier of $\theta \approx \unit[3]{mrad}$ and $\tau \approx 0$, giving rise to predominantly spatial fringes, is employed for these experiments. Typical data treated according to this process are shown in Fig.\ \ref{fig:data}.

In order to calibrate the intrinsic added second- and higher-order phase associated with the two arms of the interferometer (and specifically the dispersion of the beamsplitter and waveplate), an SSI measurement was taken with the AOPDF removed. The extracted relative higher-order phase varied by less than \unit[0.4]{rad} over the extent of the imaged spectrum. 

\section{Results}
\label{sec:results}

\begin{table}
\begin{center}
\renewcommand{\arraystretch}{1.5}
\begin{tabular}{|l|c|}
\hline
Pulse shape & $H(\omega)$ \\
\hline
Pulse delay & $\exp\sqbr{-2(\omega-\omega_0)^2/\Delta\omega^2 + i\omega \tau}$ \\
$N$-pulse train & $\sum_{n=1}^N \exp(i \omega\tau_n)$\\
Chirps & $\exp \sqbr{-2(\omega-\omega_0)^2/\Delta\omega^2 + i\br{\omega-\omega_0}^2 \phi^{(2)}/2}$\\
$\pi$-step & $\exp\left\{ i \arctan \sqbr{(\lambda-\lambda_0)/\Delta\lambda_{\textrm{step}}} \right\}$ \\
\hline
\end{tabular}
\end{center}
\caption{Transfer function, $H(\omega)$, for the pulse shapes presented within Section \ref{sec:results}, where $\omega_0 = 2\pi c/\lambda_0$ is the central angular frequency. For the pulse delay and chirped-pulse cases, a narrowed spectral bandwidth of \mbox{$\Delta\lambda = (\lambda_0^2/2\pi c)\Delta\omega = \unit[1]{nm}$} was employed. All other parameters are defined in the text.}
\label{tab:transfer-fns}
\end{table}

A series of different phase and amplitude profiles that are of broad utility within control experiments were programmed into the AOPDF. In each case, it was verified that the device applied the correct complex
transfer function $H(\omega)$, such that the input and shaped output pulses were related by $E_\textrm{out}(\omega)= H(\omega)E_\textrm{in}(\omega)$. The spatial resolution of our system enabled this verification to be performed independently at all points in the beam.

It was also possible to test systematically for any spatial or spatio-temporal distortions caused by the AOPDF. In all cases, exactly one such distortion was detected: a frequency-dependent lateral
displacement of the output proportional to the applied group delay. Pulse shapes that entailed a range of group delays across the spectral bandwidth featured a corresponding spatio-spectral coupling in the output. In all cases, the effect was consistent with a coupling speed of \unit[0.25]{mm/ps} (i.e.\ a relative lateral displacement of the shaped pulse of \unit[0.25]{mm} per picosecond shift in the AOPDF diffraction window). No other spatial or spatio-temporal distortions were detected.

The precision of the measurements was as follows. In measuring the zeroth and first-order phase components of the AOPDF transfer function, the dominant source of error was instability of the interferometer, typically \unit[0.5]{fs} over the approximately fifteen-minute durations of the data acquisition runs. For measuring higher-order phase terms, as well as the amplitude of the AOPDF transfer function, the two most significant sources of error were camera shot noise and shot-to-shot fluctuations in the UV source itself. These limited the root-mean-square precision of the phase and intensity measurements to \unit[0.2]{rad} and \unit[10]{\%} respectively. These figures apply to regions where the intensity is greater than \unit[10]{\%} of the peak. It was verified that the presence of the AOPDF did not increase the size of any phase or intensity fluctuations.

The demonstration of such spatio-temporal coupling effects --- well known and studied for the case of $4f$-line pulse shaping --- gives important information to experimentalists wishing to use AOPDFs in a control experiment. These results are presented individually below. For each case, a mathematical expression for the transfer function employed is presented in Table \ref{tab:transfer-fns}.

\subsection{Pulse delay}
\label{sec:delays}

\begin{figure}
\centering
\includegraphics [width = 0.9\columnwidth]{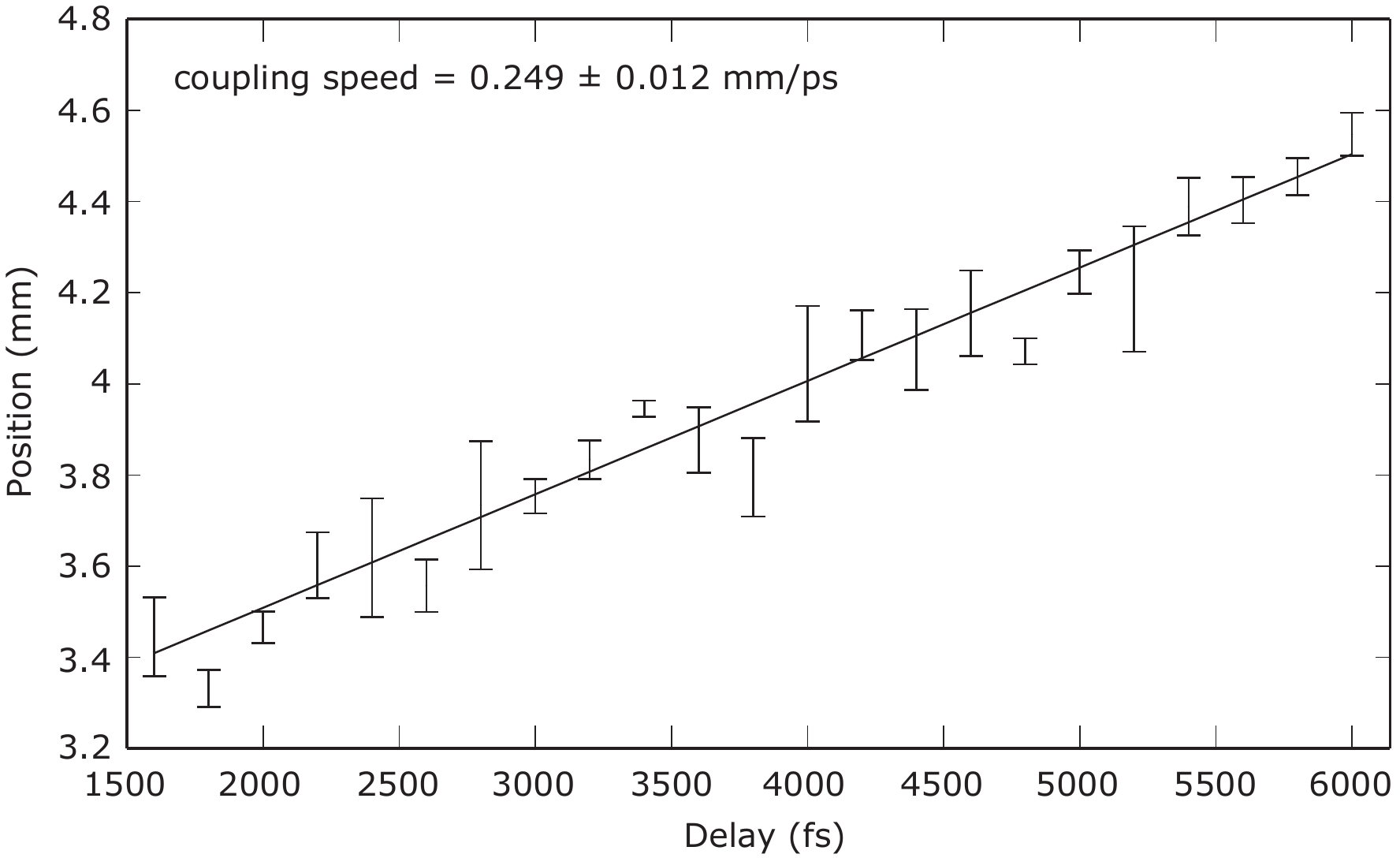}
\caption{Spatio-temporal coupling for a single optical pulse as the diffraction position within the AOPDF crystal is varied. The central beam position along the spatial axis of the spectrometer is plotted as a function of delay $\tau$. A linear dependence is observed with a best-fit gradient of \mbox{\unit[0.249 $\pm$ 0.012]{mm/ps}}.}
\label{fig:delays}
\end{figure}

For the first experiment, an acoustic wave was launched inside the AOPDF that was designed to diffract a single optical pulse within the KDP crystal. The location of the acoustic wave was scanned along the length of the crystal in order to vary the pulse delay $\tau$. The acoustic wave was tailored in order to pre-compensate for the dispersion of the crystal, and the performance of this compensation was verified via the SI measurements. The pulse spectral FWHM intensity bandwidth was also narrowed using the AOPDF to $\Delta\lambda = (\lambda_0^2/2\pi c)\Delta\omega = \unit[1]{nm}$, where $c$ is the speed of light. This reduced the length of acoustic wave required to compensate for the crystal dispersion to \unit[2]{ps}, enabling a greater range of delays to be accessed without clipping the acoustic wave on the edges of the crystal.

The measured delays were found to be in agreement with the target delays to within an error of \unit[2]{\%}. The central beam position of the diffracted pulse was observed to vary linearly with delay with a coupling speed of \mbox{\unit[0.249 $\pm$ 0.012]{mm/ps}}. Not other variation, in either amplitude or phase, was identified. The results are presented in Fig.\ \ref{fig:delays}. This behaviour was also confirmed with a direct measurement of the beam position on a CCD camera.

\subsection{Pulse train}
\label{sec:train}

\begin{figure}
\centering
\includegraphics [width = \columnwidth]{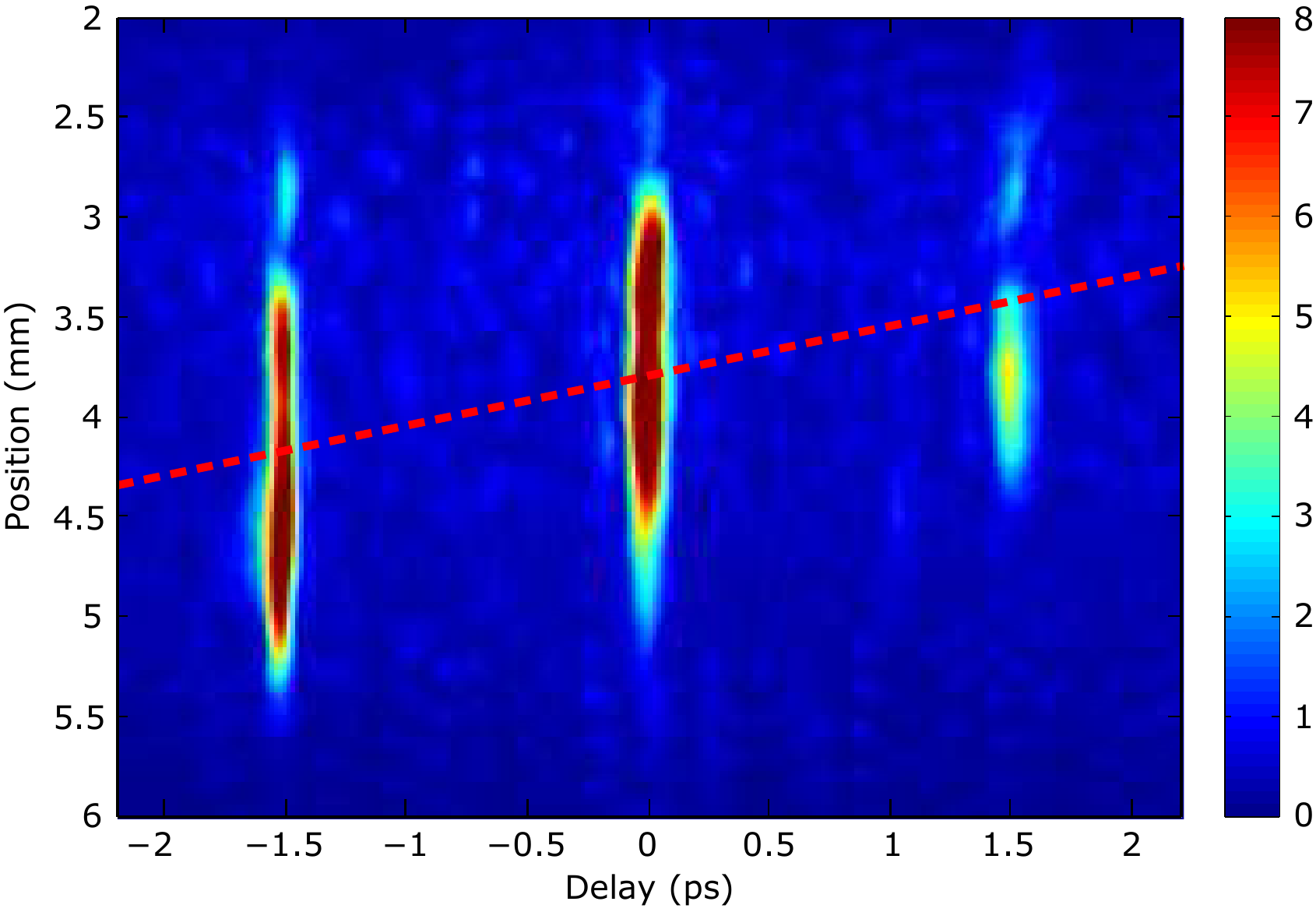}
\caption{The reconstructed spatio-temporal amplitude distribution of a train of three pulses each separated by \unit[1.5]{ps}. The reconstructed pulse train exhibits a linear spatio-temporal coupling effect that is consistent with the \unit[0.25]{mm/ps} best-fit gradient observed for the pulse delay experiments (superimposed dotted line).}
\label{fig:train}
\end{figure}

Next, various trains of pulses with zero added second- and higher-order phase were prepared, with varying numbers of pulses ranging from two to thirteen. This entailed a sequence of acoustic waves localized at different points along the length of the AOPDF crystal.

A typical reconstructed spatio-temporal intensity distribution is shown in Fig.\ \ref{fig:train} for a train of three pulses separated by \unit[1.5]{ps} (such that $N=3$, $\tau_1 = \unit[-1.5]{ps}$, $\tau_2 = \unit[0]{ps}$ and $\tau_3 = \unit[1.5]{ps}$ according to the expression of Table \ref{tab:transfer-fns}). The pulse separation was verified to within \unit[1]{\%}. In order to make the most accurate measurement possible, the full temporal window of the pulse shaper was employed. The spatio-temporal coupling subsequently resulted in a worsened alignment for the third pulse in the train, concomitantly reducing the fringe visibility. This accounts for the apparent reduction in intensity for the final pulse in Fig.\ \ref{fig:train}.

A pronounced linear spatio-temporal coupling is observed in the reconstruction. The results are quantitatively consistent with the coupling speed observed during the pulse delay experiments (Section \ref{sec:delays}), as evinced by the superimposed best-fit line with the same coupling-speed gradient of \unit[0.249]{mm/ps}.

\subsection{Chirps}
\label{sec:chirps}

\begin{figure}
\centering
\includegraphics [width = \columnwidth]{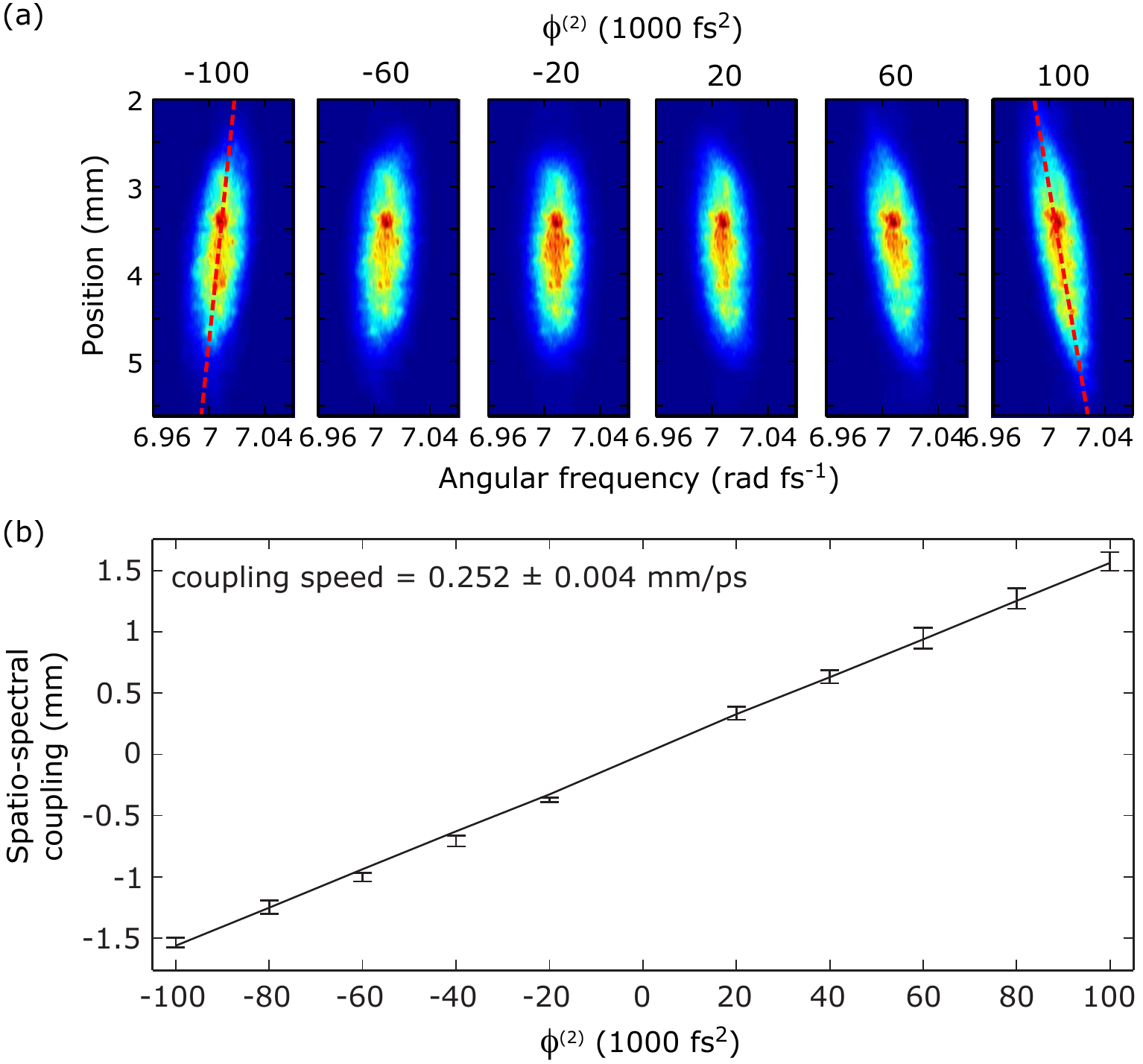}
\caption{Spatio-spectral coupling effects for a series of chirped pulses ($\phi^{(2)}$ parameters as shown). (a) The reconstructed spatio-spectral intensities of pulses of different chirps. A spatio-spectral tilt is observed that is more significant for the more strongly chirped pulses and that changes sign with the sign of the chirp. (b) The extracted spatio-spectral coupling as a function of chirp (`+') together with a calculated best-fit coupling speed of \mbox{\unit[0.252 $\pm$ 0.004]{mm/ps}} (solid line). This value is in close agreement with the measurement of Section \ref{sec:delays}; the reconstructed pulse was otherwise found to be free of further spatio-temporal coupling effects. The vertical axis shows the change in central position of the beam across the spectral bandwidth of the pulse.}
\label{fig:chirp}
\end{figure}

A range of different pulses were prepared using the AOPDF bearing different chirps --- i.e.\ the parameter $\phi^{(2)}$ in Table \ref{tab:transfer-fns}. The AOPDF temporal shaping window allowed values within the range $\unit[-100000]{fs^2} \leq \phi^{(2)} \leq \unit[100000]{fs^2}$ to be assayed. A narrowed pulse bandwidth of $\Delta\lambda = \unit[1]{nm}$ was once again employed.

The extracted $\phi^{(2)}$ second-order polynomial phase coefficients matched the programmed values to within \unit[6]{\%}. The spatio-spectral intensities are shown in Fig.\ \ref{fig:chirp}(a) for a selection of $\phi^{(2)}$ values. A spatio-spectral tilt is observed that is stronger for more strongly chirped pulses and changes sign as the sign of the chirp is reversed [see the dashed lines of \mbox{Fig.\ \ref{fig:chirp}(a)}]. This observation has important consequences for control experiments with regard to spatial alignment with the sample. Besides the spatio-spectral tilt illustrated in Fig.\ \ref{fig:chirp}(a), however, the reconstructed pulse was found to reproduce the programmed pulse with good fidelity.

The spatio-spectral tilts for a range of chirps were extracted numerically and plotted in Fig.\ \ref{fig:chirp}(b). Since spectral chirp is intrinsically a frequency-dependent group delay, the best-fit gradient of these points can be related to a group-delay--dependent displacement via the corresponding chirped-pulse temporal duration. This fit took into account an intrinsic spatio-spectral tilt present on the reference beam corresponding to a \unit[0.35]{mm} shift in beam centre across the spectral bandwidth. The best-fit coupling speed for these experiments was \mbox{\unit[0.252 $\pm$ 0.004]{mm/ps}}, in very close agreement with Sections \ref{sec:delays} and \ref{sec:train}. This demonstrates that one single underlying physical mechanism is responsible for the different spatio-temporal coupling effects.

\subsection{$\pi$-step}
\label{sec:pi-step}

\begin{figure}
\centering
\includegraphics [width = \columnwidth]{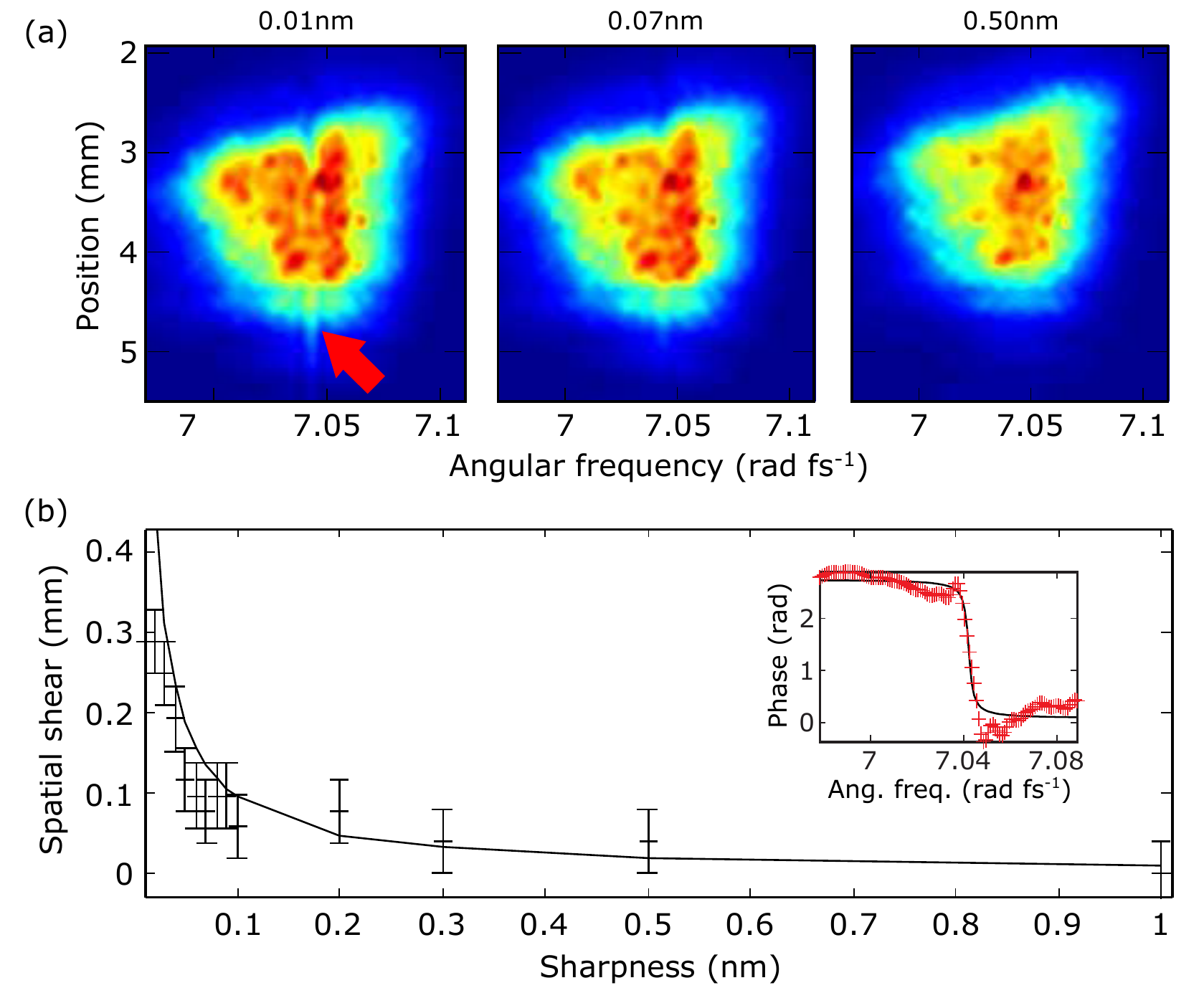}
\caption{Spatio-spectral coupling effects for $\pi$ phase-steps of varying sharpnesses. (a) The reconstructed spatio-spectral intensities of a series of $\pi$ phase-steps with sharpnesses as indicated. A spatial shift is observed at the step frequency that is more pronounced for sharper steps. (b) Observed lateral displacement of the notch (data points) together with a calculation derived from the measured group delay at the phase step and a \unit[0.25]{mm/ps} spatio-temporal coupling speed (solid line). An example spectral phase across a slice through the middle of the pulse is shown inset (crosses) together with a fit of the function in Table \ref{tab:transfer-fns} (solid line).}
\label{fig:pi}
\end{figure}

The final experiment entailed the preparation of a transform-limited pulse with a $\pi$ phase-step at its central frequency as per the expression in Table \ref{tab:transfer-fns}. Phase steps of a range of sharpnesses, $\Delta\lambda_{\textrm{step}}$, were prepared. The results are shown in Fig.\ \ref{fig:pi}(a). A typical measured spectral phase across the centre of the pulse is shown inset in \mbox{Fig.\ \ref{fig:pi}(b)}. In general, the retrieved phase matched the programmed one with regard to the parameters of Table \ref{tab:transfer-fns}. The sharpest measured step sizes were of the order on \unit[0.08]{nm}; however, this was commensurate with the resolution of the spectrometer.

An important spatio-temporal coupling effect is observed in the reconstructed spectral intensities. A local spatial displacement occurs in the spectrum at the $\pi$-step frequency, resulting in a `notch' in the reconstructed spatio-spectral intensity [see arrow in \mbox{Fig.\ \ref{fig:pi}(a)]}. The size of the notch increases with the sharpness of the phase step. This spatio-temporal coupling effect has previously only been reported in a $4f$ zero-dispersion line \cite{Dorrer1998}; this study reveals similar behaviour for an AOPDF-based device.

This notch effect may once again be reconciled with a group-velocity--dependent displacement of the beam. The steep phase gradient at the location of the $\pi$ step is equivalent to a local group-delay term in the spectral phase, with a sharper step implying a steeper gradient in the spectral phase and hence a larger group delay. A group-velocity--dependent displacement therefore shifts spectral components spanned by $\Delta\lambda_{\textrm{step}}$ by an amount dependent on the step sharpness. A related effect in pixellated SLM pulse shapers is the complete spectral hole that appears for a sharpness equal to the spectral resolution of the device \cite{Wohlleben2004}. As the step sharpness is further increased in these AOPDF experiments, the $\pi$-step group delay will eventually exceed the temporal window of the crystal (which is inversely proportional to the AOPDF spectral resolution), and a spectral hole, rather than a notch, will be formed as a consequence.

This argument is supported by the calculations presented in Fig.\ \ref{fig:pi}(b) based on these experimental data. In this figure, the notch sizes for each image within Fig.\ \ref{fig:pi}(a) were extracted and plotted as a function of step sharpness (data points). The local group-delay terms at the phase step were calculated according to $\phi^{(1)} = \frac{\partial \phi \br{\omega}}{\partial \omega} \approx \frac{\pi}{\Delta \omega}$ where $\Delta \omega_{\textrm{step}} = (2\pi c/\lambda_0^2) \Delta\lambda_{\textrm{step}}$. They were then multiplied by the \unit[0.25]{mm/ps} coupling speed previously observed (solid line), and the resultant calculation shows good agreement with experiment (solid line). Once again, the results are found to be quantitatively consistent with the same spatio-temporal coupling effect as above, reinforcing the evidence for a single underlying physical mechanism for all of these manifestations.

\section{Discussion}
\label{sec:discussion}

Section \ref{sec:results} presented a spatially resolved SSI analysis of a range of different pulse shapes: a single transform-limited pulse with a variable delay, a train of pulses, chirped pulses and pulses with a $\pi$ phase-step at the centre of their spectrum. These pulse shapes lie at the heart of many ultrafast quantum control experiments and this study represents the first complete investigation of spatio-temporal coupling effects performed for an AOPDF pulse shaper. In each case, spatio-temporal and spatio-spectral couplings were observed in the reconstructed field. Each effect was shown to be consistent with a single effect that took the form of a group-delay--dependent position of the shaped pulses, as mentioned previously \cite{Krebs2010}. Incidentally, the time-to-space mapping produced by this coupling means that the spatio-spectral intensity profile of the Dazzler output pulse resembles a spectrogram, assuming that the Dazzler input pulse is near transform-limited so that the only contribution to the group-delay in the output arises from the Dazzler itself. Furthermore, each coupling was consistent with a coupling speed of \unit[0.25]{mm/ps}. No further spatio-temporal coupling effects were identified, and the AOPDF was otherwise found to reproduce the programmed pulse shapes faithfully. In particular, no significant angular dispersion effects (as reported by B\"{o}rzs\"{o}nyi \etal \cite{Borzsonyi2010}) were found. This is to be expected since B\"{o}rzs\"{o}nyi \etal only found this to be significant at high repetition rates where the acoustic-wave energy dissipation gave rise to thermal effects.

The results above highlight the need for experimentalists to pay close attention to these coupling issues during the design of control experiments based on an AOPDF pulse shaper. Such concerns have been studied extensively for the more widespread $4f$-line shapers, with coupling speed ranging from  \unit[0.083]{mm/ps} \cite{Monmayrant2004} through \unit[0.145]{mm/ps} \cite{Wefers1996} to \unit[0.595]{mm/ps} \cite{Tanabe2002} already reported in the literature. For the $4f$-line geometry, the coupling speed $v$ is related to the available temporal shaping window $T$ and the input beam waist $\Delta x_\textrm{in}$ by $\modbr{v} = \Delta x_\textrm{in}/T$ \cite{Monmayrant2010}. The coupling speed reported here of \unit[0.25]{mm/ps} is therefore non-negligible by comparison.

It is thus apparent that a single spatio-temporal coupling mechanism within the AOPDF accounts for all the manifestations reported in Section \ref{sec:results}. In order to explain the physical nature of this group-delay--dependent displacement, it is necessary to consider a couple of effects present within the Dazzler: the birefringent and geometrical walk-off effects of the diffracted relative to the undiffracted beam, and the fact that each optical wavelength within the ultrafast pulse is diffracted at a given position in the AOPDF. These two effects combine to lead to a natural spatial chirp, with a coupling speed as quantified above.

To recapitulate, the birefringent walk-off concerns the phenomenon that the intensity distribution of a beam in an anisotropic crystal drifts away from the direction of the wave vector. The angle between the Poynting vector (which defines the direction of energy transport) and the $k$-vector is called the walk-off angle. Spatial walk-off occurs only for a beam with extraordinary polarization, which sees a refractive index $n_{\textrm{e}}$ during its propagation that depends on the angle between $\boldsymbol{k}$ and the optical axes. This angle depends on the crystal and parameters of the optical pulse; for the KDP crystal in this experiment, at \unit[268]{nm}, the walk-off angle is $\alpha \simeq \unit[32]{mrad}$. The geometrical walk-off, meanwhile, concerns the fact that during Bragg diffraction the beam is deviated by an angle corresponding to the phase-matching condition. For this experiment, this deviation is $\theta= \unit[-5.2]{mrad}$. It should be noted that both the geometric and birefringent walk-offs actually vary as a function of wavelength; however, this effect is negligible for the pulse bandwidth employed.

Thus the spatio-temporal effect can simply be seen as a shift $\delta x$ in the position of the diffracted beam that could be expressed as $\delta x =L\tan(\theta+\alpha)$, where $L$ is the distance of propagation along the extraordinary axis. The coupling speed is thus determined by $v = \delta x/T$, being a function of the walkoff-induced shift and the temporal window, rather than the input beam waist as for the case of a $4f$ line. This experiment employs a crystal of length \unit[75]{mm} such that the maximum shift is calculated as \unit[2]{mm}. Considering the fact that the temporal window available at this wavelength is $T = \unit[7.7]{ps}$, this implies an expected group-delay--dependent displacement of \mbox{\unit[$0.260 \pm 0.005$]{mm/ps}}, which is in very close agreement with our experimental measurements.

The birefringent and geometric walkoff effects are therefore confirmed as the single physical cause for the spatio-temporal coupling effect reported in the AOPDF pulse shaper. This coupling has important consequences for the application of AOPDF-shaped pulses to control experiments, since the displacement of the control pulses with a variation of pulse parameters may result in a worsened alignment with the target. One possible solution is to translate a lens before the AOPDF in order to bring the geometric plane of overlap of the spatially shifted output pulses into alignment with the gaussian focal plane \cite{Krebs2010}. Another might be to extend the walk-off compensation methods developed in non-linear optics be using a double-pass setup or a second crystal \cite{Smith1998}. It should be noted that the coupling speed depends on the parameters of the ultrafast pulses as well as the choice of crystal (indeed, the walk-off effects in TeO$_2$, which is used for AOPDFs in the IR wavelength range, are significantly less than in KDP); thus the calculation should be repeated along the lines above in order to make an informed choice of shaper in light of individual experimental tolerances for coupling effects.

\section{Conclusion}
\label{sec:conclusion}

In this paper, we have presented a systematic study of spatio-temporal coupling in an AOPDF pulse shaper that operates at UV wavelength ranges via spatially resolved phase and amplitude analysis of the shaped pulses. Such coupling effects have been widely studied for $4f$ zero dispersion lines due to the importance of the ramifications for control experiments. The AOPDF is an increasingly popular alternative shaping device thanks to its versatility, compactness, ease of alignment and wide wavelength range. Until now, however, its spatio-temporal coupling effects have not been comprehensively studied for a range of complex pulse shapes of interest to the control community.

We have discovered that there is one single significant effect at kilohertz repetition rates: a group-delay--dependent displacement of the shaped output. Further to this one effect, the AOPDF was found to produce faithfully the desired pulse shape. This coupling effect was manifested differently in the measured pulse depending on the class of pulse shape employed; however, in each case the coupling effect may be described by the same mechanism with consistent quantitative agreement. We have explained the physical origin of this mechanism and have shown excellent agreement between its calculated and measured values. Finally, we have identified some approaches that may allow the impact of this spatio-temporal coupling to be minimized during applications to control experiments.

\begin{acknowledgements}
The authors are grateful to N.\ Forget and A.\ Wyatt for useful discussions as well as to E.\ Baynard and S.\ Faure for technical assistance. This work was supported by the Marie Curie Initial Training Network (grant no.\ CA-ITN-214962-FASTQUAST), EPSRC grants EP/H000178/1 and EP/G067694/1, and Alliance (PHC/British Council).
\end{acknowledgements}

\end{document}